\documentclass[peerreview]{IEEEtaes}

\usepackage[utf8]{inputenc}
\usepackage[T1]{fontenc}

\usepackage{cite}

\usepackage[colorlinks,urlcolor=blue,linkcolor=black,citecolor=black]{hyperref}

\usepackage{color,array,amsthm}
\usepackage{graphicx}

\usepackage{comment}
\usepackage{bm}
\usepackage{bbold}
\usepackage{mathtools}
\usepackage{bbm}
\usepackage{dsfont}

\usepackage{siunitx}
\usepackage{latexsym}
\usepackage{amsmath}
\usepackage{amsfonts}
\usepackage{amssymb}
\usepackage[capitalise,noabbrev]{cleveref}
\usepackage{multirow}
\usepackage{makecell}
\usepackage{xcolor,colortbl}

\usepackage{hhline}
\usepackage[skip=3pt]{caption}

\newlength\imagewidth
\newlength\imagescale

\definecolor{Gray}{gray}{0.85}

\usepackage{subcaption}
\usepackage{hyperref}
\usepackage{listings}
\usepackage{booktabs}
\usepackage{placeins}
\usepackage{float}
\usepackage{array}
\newcolumntype{?}{!{\vrule width 1pt}}

\usepackage{pgfplots}
\pgfplotsset{compat=newest}
\usetikzlibrary{plotmarks}
\usetikzlibrary{arrows.meta}
\usepgfplotslibrary{patchplots}
\usepackage{grffile}
\pgfplotsset{plot coordinates/math parser=false}
\newlength\figureheight
\newlength\figurewidth
\newcommand{\virg}[1]{``#1''}

\jvol{XX}
\jnum{XX}
\jmonth{XXXXX}
\paper{1234567}
\pubyear{2020}
\doiinfo{TAES.2020.Doi Number}

\begin{document}

\sptitle{Correspondence}
\title{Enhancement of a state-of-\\the-art RL-based detection \\ algorithm for Massive MIMO \\radars}

\author{FRANCESCO LISI}
\member{Student Member, IEEE}
\affil{Dipartimento di Ingegneria dell’Informazione, Università di Pisa, Italy} 

\author{STEFANO FORTUNATI}
\member{Senior Member, IEEE}
\affil{Université Paris-Saclay, CNRS, CentraleSupélec, L2S, Gif-sur-Yvette

\&  DR2I-IPSA, Ivry-sur-Seine, France} 

\author{MARIA SABRINA GRECO}
\member{Fellow, IEEE}
\affil{Dipartimento di Ingegneria dell’Informazione, Università di Pisa, Italy}

\author{FULVIO GINI}
\member{Fellow, IEEE}
\affil{Dipartimento di Ingegneria dell’Informazione, Università di Pisa, Italy}

\receiveddate{The work of Francesco Lisi was supported by the Laboratoire des signaux et systèmes (L2S). The work of M.S. Greco and F. Gini has been partially supported by the Italian Ministry of Education and Research (MIUR) in the framework of the CrossLab project (Departments of Excellence) of the University of Pisa, laboratory of Industrial Internet of Things (IIoT).}

\corresp{{\itshape (Corresponding author: F. Lisi)}.}

\authoraddress{F. Lisi, M. S. Greco and F. Gini are with Dipartimento di Ingegneria dell’Informazione, Università di Pisa, Italy (e-mail: \href{mailto:francesco.lisi@phd.unipi.it}{francesco.lisi@phd.unipi.it}, \href{mailto:maria.greco@unipi.it}{maria.greco@unipi.it}, \href{mailto:fulvio.gini@unipi.it}{fulvio.gini@unipi.it}). S. Fortunati is with Université Paris-Saclay, CNRS, CentraleSupélec, Laboratoire des signaux et systèmes, 91190, Gif-sur-Yvette \& DR2I-IPSA, 94200, Ivry-sur-Seine, France (e-mail: \href{mailto:stefano.fortunati@centralesupelec.fr}{stefano.fortunati@centralesupelec.fr}).}

\editor{The code related to this work can be found at \mbox{\url{https://github.com/lisifra96/Improved_RL_algorithm_mMIMO_radar}} }
\supplementary{Color versions of one or more of the figures in this article are available online at \url{http://ieeexplore.ieee.org}.}

\markboth{CORRESPONDENCE}{}
\maketitle

\begin{abstract}
In the present work, a reinforcement learning (RL) based \textit{adaptive} algorithm to optimise the transmit beampattern for a co-located massive MIMO radar is presented. Under the massive MIMO regime, a robust Wald-type detector, able to guarantee certain detection performances under a wide range of practical disturbance models, has been recently proposed. Furthermore, an RL/cognitive methodology has been exploited to improve the detection performance by learning and interacting with the surrounding unknown environment. Building upon previous findings, we develop here a fully adaptive and \textit{data-driven} scheme for the selection of the hyper-parameters involved in the RL algorithm. Such an adaptive selection makes the Wald-RL-based detector independent of any ad-hoc, and potentially sub-optimal, manual tuning of the hyper-parameters. Simulation results show the effectiveness of the proposed scheme in harsh scenarios with strong clutter and low SNR values.  
\end{abstract}


\section{INTRODUCTION}
The concept of Cognitive Radar (CR) has been firstly introduced by Haykin \cite{haykin2006cognitive}. The need of adaptation to the changes in the environment is intrinsic in the radar detection problem due to the presence of multiple sources of non-stationarity, such as variations in the clutter statistics over time or changes in the target scenario. Unlike communication systems where the transmitter and receiver are physically separated, in a monostatic radar system the transmitter and the receiver are in the same position, allowing the latter to easily transmit information to the former. MIMO radar systems can be divided into two fundamentals categories: widely separated radars and co-located radars. As \cite{haimovich2007mimo} suggests, a radar with widely separated antennas can exploit the spatial diversity of the target radar cross section to obtain a diversity gain similar to the one in communications. In a co-located MIMO radar, the antennas are closely spaced and each element of the array transmits a different probing signal, contrary to standard phased array where all the elements transmit the same waveform with variable amplitude and phase. In \cite{li2007mimo} the authors discuss the advantages of parameter identifiability led by co-located MIMO radars. Following the recent works \cite{fortunati2020massive,ahmed2021reinforcement},
in this correspondence, we focus our attention on co-located massive MIMO radars exploiting RL-based techniques \cite{sutton2018reinforcement,mohri2018foundations} to implement the \textit{cognition loop} \cite{greco2018cognitive}. 
Specifically, in \cite{fortunati2020massive} the authors proved that, if the number of virtual spatial antenna channels is high enough (massive MIMO regime), a robust Wald-type test can be derived to guarantee the constant false alarm rate (CFAR) property under a wide variety of disturbance models using a single snapshot. Then, in \cite{ahmed2021reinforcement}, a reinforcement learning (RL) approach is proposed to optimise the transmission beampattern of a co-located massive MIMO radar exploiting the robust Wald-type detector in \cite{fortunati2020massive}. 

Two original contributions are proposed in this correspondence. The first one is the introduction of two new policies, called \textit{quasi $\varepsilon$-greedy policy} and \textit{quasi $\varepsilon$-greedy policy with target recovery}. The former improves the performance of the system by forcing the RL algorithm to focus its power in a number of angular bins greater or equal to the number of detected targets; in addition, the latter exploits a mechanism that allows the system to recover a missed target more quickly. The second original contribution consists of an adaptive algorithm that selects the SARSA algorithm hyper-parameters ($\varepsilon$ and $\alpha$) based on the received signal. This allows the system to adapt to the variations of the surrounding environment and to achieve better detection capability both in stationary and non-stationary scenarios. Afterwards, the RL-based algorithm proposed in \cite{ahmed2021reinforcement} with the quasi $\varepsilon$-greedy policy with target recovery and the adaptive selection of $\varepsilon$ and $\alpha$ is tested in various simulation scenarios. The results confirmed that the updated algorithm achieves higher detection performance, compared to the one in \cite{ahmed2021reinforcement}. Both our algorithm and the one proposed in \cite{ahmed2021reinforcement} exploit the robust Wald-type detector described in \cite{fortunati2020massive}.

In \cref{sec:THE CFAR DETECTION ALGORITHM} we describe the detection problem and the main properties of the robust Wald-type detector derived in \cite{fortunati2020massive}. \cref{sec:RL-BASED ALGORITHM} provides the background on the RL-based detector proposed in \cite{ahmed2021reinforcement}. In \cref{sec: policy improvement} two variations of the $\varepsilon$-greedy policy are proposed, while an adaptive algorithm to select the $\varepsilon$ and $\alpha$ parameters is described in \cref{sec:adaptive algorithms}. In \cref{sec:SIMULATION RESULTS} we validate the proposed contributions via simulation. The simulated scenarios are described in Appendix \ref{appendix}. Finally, conclusions are drawn in \cref{sec:CONCLUSION}.  

\textbf{NOTATIONS:} In this paper we use upper case letters $\bm{A}$ and lower case ones $\bm{a}$ to denote matrices and vectors respectively. $(\cdot)^T$ and $(\cdot)^H$ denote a matrix transpose and conjugate transpose respectively, while $(\cdot)^*$ denotes the conjugate operator. $\bm{I}_{N}$ denotes the $N \times N$ identity matrix, while $\bm{0}_{N}$ denotes an all zeros $N \times N$ matrix. $E\{\cdot\}$ denotes the statistical expectation. The Kronecker product is represented by $\otimes$. A closed interval of numbers between $a$ and $b$ is denoted by $[a,b]$, while a set containing only $a$ and $b$ is denoted by $\{a,b\}$. The absolute value is represented by $|\cdot|$. The relative complement of set $\mathcal{A}$ with respect to set $\mathcal{B}$ is denoted as $\mathcal{B}\backslash\mathcal{A}$.


\section{THE CFAR DETECTION ALGORITHM}
\label{sec:THE CFAR DETECTION ALGORITHM}

Consider a co-located MIMO radar with $N_T$ transmit and $N_R$ receive antennas. Both transmit and receive arrays are uniform linear arrays (ULA) with inter-element distance $d=\lambda/2$. The transmitted signal vector can be expressed as $\bm{s}(t)= \bm{W}\bm{\Phi}(t)$ where $\bm{W} \in \mathbb{C}^{N_T \times N_T}$ is a weighting matrix and $\bm{\Phi}(t)\in \mathbb{C}^{N_T}$ is a vector containing a set of orthonormal signals. After sampling the correct range-doppler bin at the output of the matched filter the received signal can be expressed as \cite{fortunati2020massive}
\begin{equation}
	\label{eq:receiver output matched filter(vector)}
	\bm{y}=\alpha \bm{h}+\bm{c} \in \mathbb{C}^{N \times 1},
\end{equation}
with $N=N_T N_R$ and $\bm{h}= \bm{W}^T \bm{a}_T(\nu_0) \otimes \bm{a}_R(\nu_0)$. Finally, $\bm{a}_T(\nu_0)$ and $\bm{a}_R(\nu_0)$ are the transmit and receive steering vectors that depend on the spatial frequency $\nu_0=(d/\lambda)\sin(\theta_0)$ where $\theta_0$ is the target angle of arrival (in a given reference frame).



Assuming that the received signal is processed at each time instant $k$ by a bank of $L$ spatial filters tuned to a fixed grid of angular bins $\Theta=\{\theta_l\}_{l=1}^L$, the detection problem can be formulated as the following hypothesis testing problem \cite{ahmed2021reinforcement}
\begin{equation}
	\label{eq:hypothesis test formulation}
	\begin{split}
		&\mathcal{H}_0: \bm{y}_{k,l}=\bm{c}_{k,l}, \\
		&\mathcal{H}_1: \bm{y}_{k,l}=\alpha_{k,l} \bm{h}_{k,l}+\bm{c}_{k,l}.
	\end{split}
\end{equation}

In the following, we assume that the grid is chosen in order to uniformly span the spatial frequency interval $[-0.5,0.5]$. 
\\ As a test statistic, the robust Wald-type detector $\Lambda_{k,l}$ is adopted \cite{fortunati2020massive,TYPOS}:
\begin{equation}
	\label{eq:WT statistic (time-angle)}
	\Lambda_{k,l}=\frac{2\vert\bm{h}_{k,l}^H\bm{y}_{k,l}\vert^2}{\bm{h}_{k,l}^H\hat{\bm{\Gamma}}_{k,l}\bm{h}_{k,l}},
\end{equation}
where $\hat{\bm{\Gamma}}_{k,l}$ is an estimate of the covariance matrix of the noise vector $\bm{c}_{k,l}$. The detector discriminates between $\mathcal{H}_0$ and $\mathcal{H}_1$ by comparing the statistic with a threshold $\lambda$ in each angular bin.
When $N\rightarrow\infty$, the statistic satisfies the CFAR property, i.e. the Probability of False Alarm ($P_{FA}$) is constant and can be selected by choosing the threshold as $\lambda=-2 \, \text{ln}\left({P}_{FA}\right)$, under a wide rage of practical disturbance processes with unknown statistical characterisation. For a $P_{FA}$ equal to $10^{-4}$, if $N\geq10^4$ the MIMO radar is assumed to operate in the massive MIMO regime and the previous property is satisfied \cite{fortunati2020massive}.


\section{RL-BASED ALGORITHM}
\label{sec:RL-BASED ALGORITHM}

After having ensured the CFARness of the detection scheme through the Wald-type detector in \eqref{eq:WT statistic (time-angle)}, the work \cite{ahmed2021reinforcement} focused on how to exploit the large degrees of freedom, offered by a massive MIMO radar, to maximise the Probability of Detection ($P_D$). This can be achieved by properly selecting the weighting matrix $\bm{W}_k$ to shape the transmit beampattern $\bm{a}_{T}^T(\nu) \bm{W}_k\bm{W}_k^H \bm{a}_{T}^*(\nu)$ \cite{fuhrmann2008transmit}. In order to do so, an RL-based algorithm has been proposed to allow the radar focusing the power in the angular bins associated to the targets' angular position \cite{ahmed2021reinforcement}. In the following, a concise summary of the main definitions and results obtained in \cite{ahmed2021reinforcement} will be provided since they represent the starting point of the original developments proposed in this correspondence. Let us start by introducing the Markov Decision Process (MDP) characterising the learning/detection task at hand. For the sake of clarity, it is worth mentioning that the following definitions of state, action and reward, associated to the considered MDP, are slightly different with respect to the ones introduced in \cite{ahmed2021reinforcement}. Even though the essence remains exactly the same, this new formulation is more precise and self-consistent.


\subsection{The set of the states}
\label{sec:State definition}

The state space of our MDP is denoted as \mbox{$\mathcal{S} \triangleq \big\{s^{(i)}\big\}_{i=0}^{K}$}, where $K<\infty$ is the maximum number of detectable target. 
 
The state of the system at time instant $k$ is defined as $s_k=s^{(i_k)}$ with \cite{ahmed2021reinforcement}
\begin{equation}
	\label{State index definition}
	 i_k \triangleq \min \bigg\{\sum\limits_{l=0}^{L-1}\bar{\Lambda}_{k,l} ,K \bigg\},
\end{equation}

where $\bar{\Lambda}_{k,l} \triangleq u(\Lambda_{k,l}-\lambda)$ and $u(\cdot)$ corresponds to the Heaviside step function \footnote{In this correspondence the notation $(\cdot)_{k,l}$ is used, unlike the one $(\cdot)_l^k$ used in \cite{ahmed2021reinforcement} .}. 

\subsection{The set of the actions}
\label{sec:Action definition}

The action set of the MDP is indicated as \mbox{$\mathcal{A} \triangleq \big\{a^{(j)}\big\}_{j=0}^{K}$}. If the action $a^{(j_k)}$ is selected at time instant $k$, then the beamforming algorithm focuses its power in the $j_k$ angular bins with the highest decision statistic.

Let $\big\{l_k^{(n)}\big\}_{n=1}^{L}$ be the set containing the angular bin's indexes sorted in descending order with respect to the decision statistic at time instant $k$ (\mbox{$\Lambda_{k,l_k^{(1)}}\geq ... \geq \Lambda_{k,l_k^{(L)}}$}), we define the set 
\begin{equation}
	\label{Omega set definition}
	\Omega_k \triangleq 
	\begin{cases}
		\emptyset \; , & j_k=0
		\\ \{l_k^{(n)}\}_{n=1}^{j_k} \; , & j_k\neq0
	\end{cases},
\end{equation}
containing the indexes of the angular bins associated to the $j_k$ highest decision statistics.
\\ The system chooses the matrix $\bm{W}_k$ according to \cite{ahmed2021reinforcement,wang2018reinforcement}
\begin{equation}
	\label{W optimisation problem}
	\bm{W}_k=
	\begin{cases}	
		\bm{W}_{ort}\triangleq \sqrt{\frac{P_{max}}{N_T}}\cdot\bm{I}_{N_T} \; , & \Omega_k=\emptyset \\
		\tiny{\left\lbrace \begin{split}
			&\text{arg}\max\limits_{\bm{W}}\min\limits_{l\in\Omega_k} &\bm{a}_{T}^T(\nu_l) \bm{WW}^H \bm{a}_{T}^*(\nu_l)  \\
			&\qquad\text{subject to } &\text{tr}\big\{\bm{WW}^H\big\}\leq P_{max}
		\end{split}\right.} 	\; , & \Omega_k\neq\emptyset
	\end{cases}
\end{equation} 
where $P_{max}$ is the maximum transmitted power.


\subsection{The reward function}
\label{sec:Reward definition}

Let's define the set $\Psi_k \triangleq \big\{l_k^{(n)}\big\}_{n=1}^K$, where $\big\{l_k^{(n)}\big\}_{n=1}^{L}$ is the same set as in \cref{sec:Action definition}, and the sets
\begin{equation}
	\label{phi set}
	\Phi_k \triangleq 
	\begin{cases}
		\emptyset \; , & i_k=0 \\
		\big\{l_k^{(n)}\big\}_{n=1}^{i_k} \; , & i_k \neq 0
	\end{cases},
\end{equation}
and $\bar\Phi_k \triangleq \Psi_k \backslash \Phi_k$, where $i_k$ is defined in \eqref{State index definition}.

The reward is defined as \cite{ahmed2021reinforcement}
\begin{equation}
	\label{reward definition}
	r_k \triangleq \sum\limits_{l\in \Phi_k}\hat P_{D,k,l} - \sum\limits_{l\in \bar\Phi_k}\hat P_{D,k,l}
\end{equation}
where $\hat P_{D,k,l}$ is an estimate of the $P_D$ of a target located in the $l^{th}$ angular bin at the $k^{th}$ time instant. 


\subsection{SARSA algorithm}

The goal of RL algorithms is to find the optimal policy $\pi^*$, i.e. the one that maximises the \textit{state value function} $V_{\pi}(s) \triangleq E_{\pi} \left \{ \sum\limits_{h=0}^{+ \infty} \gamma^{h} r_{k+h+1}  \big \vert s_k=s \right \}$, \mbox{$\forall s \in \mathcal{S}$}, where \mbox{$\gamma \in [0,1]$} is a damping factor. By defining the \textit{state-action value function} associated to the policy $\pi$ as \mbox{$Q(s,a) \triangleq E_{\pi} \left \{ \sum\limits_{h=0}^{+ \infty} \gamma^{h} r_{k+h+1} \big \vert (s_k=s) \cap (a_k=a) \right \}$}, it can be proved that the greedy action associated to the state $s$ is equal to \mbox{$\pi^*(s)=\mathrm{arg}\max_{a' \in \mathcal{A}} Q_{\pi^*}(s,a')$} \cite{mohri2018foundations, sutton2018reinforcement}. The SARSA algorithm, named after the update rule sequence \textit{State-Action-Reward-State-Action}, allows the system to recursively compute the $\bm{Q}$ matrix associated to the optimal policy when the dynamics of the environment are unknown. 

The algorithm proposed in \cite{ahmed2021reinforcement} consists of setting the initial values $s_0=s^{(0)}$, $a_0=a^{(0)}$ and $\bm{Q}_0=\bm{0}_{K+1}$ and proceeding by computing the new state $s_{k+1}$ and reward $r_{k+1}$, selecting a new action $a_{k+1}$ following the $\varepsilon$-greedy policy and then updating $\bm{Q}$ as 
\begin{equation}
	\label{eq:SARSA Q update rule}
	\begin{split}
		Q_{k+1}&(s_k,a_k)=Q_{k}(s_k,a_k)+\\
		&+\alpha_{k+1}(r_{k+1}+\gamma Q_{k}(s_{k+1},a_{k+1})-Q_{k}(s_k,a_k)).	
	\end{split}
\end{equation} 

Once the action has been selected the algorithm computes the $\bm{W}_{k+1}$ matrix following \eqref{W optimisation problem} and then the system transmits the new pulses. The \textit{learning rate} $\alpha_k \in (0,1)$ is the weight given to the new information with respect to the old one.


\subsection{$\varepsilon$-greedy policy}

To ensure the convergence of the SARSA algorithm, the new action $a_{k+1}$ must be selected according to a policy that guarantees that each state-action pair is visited infinitely many times. The $\varepsilon$-greedy policy is one of the most used ones in the RL-literature \cite{mohri2018foundations, sutton2018reinforcement}. The new action is selected as 
\begin{equation}
	\label{eq:epsilon greedy policy}
	\pi^{(1)}_k(s_k)=
	\begin{cases}
		a_{k}^{(greedy)}, \quad & \text{w.p.} \; 1-\varepsilon \\
		\mathbb{U} \left\{\mathcal{A}-\{a_{k}^{(greedy)}\} \right\}, \quad & \text{w.p.} \; \varepsilon
	\end{cases},
\end{equation}
where \virg{w.p.} stands for \textit{with probability}, $a_{k}^{(greedy)}= \mathrm{arg}\max_{a' \in \mathcal{A}} Q_{k}(s_{k},a')$ and $\mathbb{U}\{\mathcal{C}\}$ denotes a function that selects randomly one of the elements in the set $\mathcal{C}$ with uniform probability. The $\varepsilon$ parameter controls the \textit{exploration-exploitation tradeoff}: low $\varepsilon$ values correspond to a system that chooses the greedy action most of the time (exploitation), while high values to one that selects a random action more frequently (exploration).

\section{Policy improvement}
\label{sec: policy improvement}

This section and the following \cref{sec:adaptive algorithms} present the original contribution of this correspondence.
We start by providing two improved versions of the $\varepsilon$-greedy policy. 


\subsection{Quasi $\boldsymbol{\varepsilon}$-greedy policy}
\label{sec:Quasi epsilon-greedy policy}

The problem arising from choosing the $\varepsilon$-greedy policy in our specific application can be easily explained with an example.

\textit{Example 1}. Consider a static scenario with two targets as in Scenario 1 (see Appendix \ref{appendix}). Suppose that the system performs the optimal action at the $k^{th}$ iteration, i.e.  selects action $a_k=a^{(2)}$ and the set $\Omega_k$, defined in \eqref{Omega set definition}, contains the indices associated to the actual position of the two targets. If the system chooses the random action $a_{k+1}=a^{(1)}$ at the next time instant and focuses all the power in only one of the two targets, with high probability it will end up in $s_{k+2}=s^{(1)}$. At this point, even if the system selects the action $a_{k+2}=a^{(2)}$, the set $\Omega_{k+2}$ isn't guaranteed to contain the indexes of the two targets. The system will stay in state $s^{(1)}$ until the missed target will be in the set $\Omega_{k+m}$.

To overcome this issue a new policy is proposed here. We define \textit{the quasi $\varepsilon$-greedy policy} as:
\begin{equation}
\label{quasi epsilon-greedy policy definition}
	\pi_k^{(2)}(s_k)=
	\begin{dcases}
		\mathbb{U} \big\{ \mathcal{A}'(s_k)-\{a_k^{(greedy)}\} \big\} \; , & \text{w.p.}\quad \varepsilon \\
		a_k^{(greedy)}  \; , & \text{w.p.} \quad 1-\varepsilon
	\end{dcases}
\end{equation} 
where $\mathcal{A}'(s^{(i)}) \triangleq \{a^{(j)},j=i,...,K\}$. If the quasi $\varepsilon$-greedy policy is combined with $\bm{Q}_0=\bm{I}_{K+1}$ then, when the system is in state $s^{(i)}$, it can't focus its power in a number of angular bins less than $i$, that corresponds to the number of detected targets if these aren't more than $K$.
	

\subsection{Quasi $\boldsymbol{\varepsilon}$-greedy policy with target recovery}
\label{sec:Quasi epsilon-greedy policy with target recovery}
 
 Although the quasi $\varepsilon$-greedy policy solved one of the main issues encountered with the $\varepsilon$-greedy policy, it doesn't solve the \textit{target loss problem} described in the following example.
 
 \textit{Example 2}. Consider a scenario with two targets in the radar scene, like in Scenario 1 (see Appendix \ref{appendix}). Suppose that, at a given time instant $k$, the system is in state $s^{(2)}$, having detected both targets, and randomly selects action $a^{(5)}$. Although two of the five angular bins in which the system focuses its power correspond to the position of the targets, the system might lose one of them if its SNR is low and end up in state $s^{(1)}$. Now, with high probability, the two highest decision statistics still correspond to the angular position of the targets. Consequently, if the system selects action $a^{(2)}$, it recovers the lost target and gets back in state $s^{(2)}$. On the other hand, since the system is in state $s^{(1)}$, it might happen that it selects action $a^{(1)}$ randomly or because this is the greedy action associated to state $s^{(1)}$, especially in the early phases when the system hasn't figured out the scene yet. In this case the system focuses all its power in the direction of the detected target and the decision statistic of the missed one drops. Now it will take a long time for the system to recover it.
 
 To solve this \virg{missed target} problem, we define a new policy as:
 \begin{equation}
 	\label{quasi epsilon-greedy policy with target recovery}
 	\pi_k^{(3)}(s_k,s_{k-1}) \triangleq
 	\begin{cases}
 		 \mathrm{arg}\max\limits_{a \in \mathcal{A}}Q_{k}(s_{k-1},a) \; , & i_k<i_{k-1} \\
 		\pi_k^{(2)}(s_k) \; , & i_k \geq i_{k-1}.
 	\end{cases}
 \end{equation}
 
In words, this policy can be motivated as follows: when the system detects an higher or equal number of targets, it chooses the new action according to the quasi $\varepsilon$-greedy policy. On the contrary, if the number of detected targets is smaller than the one at the previous iteration the system tries to recover them as soon as possible by choosing the greedy action associated to the state at the previous time instant $s_{k-1}$. 


\section{Adaptive selection of $\boldsymbol{\varepsilon}$ and $\boldsymbol{\alpha}$}
\label{sec:adaptive algorithms}

The $\varepsilon$ and $\alpha$ hyper-parameters are both kept constant in the algorithm proposed in \cite{ahmed2021reinforcement}. Although this is one of the most popular choices for non-stationary environments, it has two major drawbacks. Firstly, in order to use the algorithm, the user must select them in advance and this operation requires a certain knowledge of both the problem and algorithm, thus limiting its usability. Secondly, the parameters' optimal value changes over time, both in stationary and non-stationary environments. To overcome these problems we propose an adaptive algorithm that selects both the parameters based on the collected data.

Let $r_k$ be the reward at time instant $k$ and $d_k$ the sequence defined as
\begin{equation}
\label{reward difference sequence definition}
	d_k \triangleq 
	\begin{cases}
		r_1 &, \; k=1\\
		r_k-r_{k-1} &, \; k \neq 1\\
	\end{cases}.
\end{equation}
Then, $\varepsilon$ and $\alpha$ can be updated, at the $k^{th}$ iteration, according to the following strategy:
\begin{equation}
\label{epsilon and alpha update equation}
	x_{k+1}=
	\begin{cases}
		\max\{c_1 \cdot x_k \;,\; x_{min} \}, \quad &|d_k|<\eta_1 \\
		\min\{c_2 \cdot x_k \; , \; x_{max} \}, \quad \eta_1<&|d_k|<\eta_2 \\
		x_{max}, \quad &|d_k|>\eta_2 \\
	\end{cases}
\end{equation}
where $x$ corresponds to $\varepsilon$ or $\alpha$, $c_1\in(0,1)$ and $c_2\in (1,+ \infty )$. The initial value of $x$ is set to $x_0=x_{max}$. \mbox{$c_1$, $c_2$, $\eta_1$ and $\eta_2$} are constants, but differ for the $\varepsilon$ and $\alpha$ algorithm. \cref{table:Adaptive epsilon and alpha algorithm parameters} lists all the constants' values. $x_{k+1}$ is not updated if the system was in exploration mode in the two previous time instants, i.e. $k-1$ and $k$. If the system selects a random action at time instant $k$, the reward $r_{k+1}$ may drop due to a bad choice of the action causing \mbox{$\vert d_{k+1} \vert = \vert r_{k+1}-r_{k} \vert$} to surpass $\eta_1$ even though the scenario hasn't changed. If the system then chooses the correct action at time $k+1$, the reward $r_{k+2}$ rises back to a value around $r_k$, but \mbox{$\vert d_{k+2} \vert = \vert r_{k+2}-r_{k+1} \vert$} is likely to be over $\eta_1$ due to the low value of $r_{k+1}$.

The thresholds $\eta_1$ and $\eta_2$ depend on the definition of the reward. For the one defined in \eqref{reward definition}, if the system misses one target or detects a new one the reward variation will be around 1 depending on the target SNR. A good choice for $\eta_1$ is 0.5. Moreover, $\eta_2$ should be chosen high enough to guarantee that the value is reset only when a sudden change in the scenario happens: some empirical tests suggested that $\eta_2 \geq 1.8$ meets this specification.

\begin{table}[t]
\centering
\def\arraystretch{1.0}
\caption{Adaptive $\varepsilon$ and $\alpha$ algorithm parameters. $\varepsilon_{min}$ $(\alpha_{min})$ and $\varepsilon_{max}$ $(\alpha_{max})$ are the endpoints of the interval to which $\varepsilon$ $(\alpha)$ belongs; $c_1$ and $c_2$ correspond to the multiplicative decrease and increase factors; $\eta_1$ and $\eta_2$ are the lower and upper threshold respectively.}
\label{table:Adaptive epsilon and alpha algorithm parameters}
	\begin{tabular}{*{7}{|>{\centering\arraybackslash} p{7mm}}|} 
 	\hline
 	$x$ & $x_{min}$ & $x_{max}$ & $c_1$  & $c_2$ & $\eta_1$ & $\eta_2$\\ 
 	\hline
 	$\varepsilon$ & 0.1 & 0.8 & 0.8 & 2  & 0.5 & 1.8\\ 
	\hline
 	$\alpha$ & 0.2 & 0.6 & 0.9 & 2.5  & 0.5 & 1.8\\ 
	\hline	
	\end{tabular}
\end{table}


\section{SIMULATION RESULTS}
\label{sec:SIMULATION RESULTS}

In this section the performances of the two policies and the adaptive algorithm previously described are validated via  simulations in Scenarios 1 and 2 (see Appendix \ref{appendix}). Then the RL-based cognitive algorithm (RL-C) with the adaptive selection of $\varepsilon$ and $\alpha$ and the quasi $\varepsilon$-greedy policy with target recovery is compared with the orthogonal algorithm and a \virg{non RL-based} cognitive algorithm (NRL-C) in two non-stationary scenarios (3 and 4). The orthogonal algorithm is the one with an omnidirectional beampattern, while the NRL-C algorithm focuses the power in the angular bins where a detection occurred. As performance benchmark, we plot an upper bound on the $P_D$ obtained by a clairvoyant beamformer that focuses its power in the exact (and generally unknown) direction of the targets. Finally, the proposed RL-C algorithm is compared with the one in \cite{ahmed2021reinforcement} in a stationary scenario. Due to a lack of space only the most relevant figures are shown, but the interested reader can obtain all the simulations' results using the code available via the link in the first page of the correspondence. \cref{table:simulation parameters} lists all the parameters' values used in the simulations.

\begin{table}[t]
\caption{Simulation parameters. $L$ is the number of spatial filters in the receiver and $K$ is the maximum number of detectable targets; $\gamma$ is the damping factor of the cumulative reward and $\bm{Q}_0$ is the initial value of the \textit{state-action value} matrix.}
\label{table:simulation parameters}
\centering
\def\arraystretch{1.2}
	\begin{tabular}{*{8}{|>{\centering\arraybackslash} p{6mm}}|} 
 	\hline
 	$P_{FA}$ & $N_T$ & $N_R$ & L  & K  & $P_{max}$ & $\gamma$ & $\bm{Q}_0$\\ 
 	\hline
 	$10^{-4}$ & 100 & 100 & 20 & 5  & 1 & 0.8 & $\bm{I}_{K+1}$\\ 
	\hline
	\end{tabular}
\end{table}

\begin{figure}[t!]
  \centering
  \input{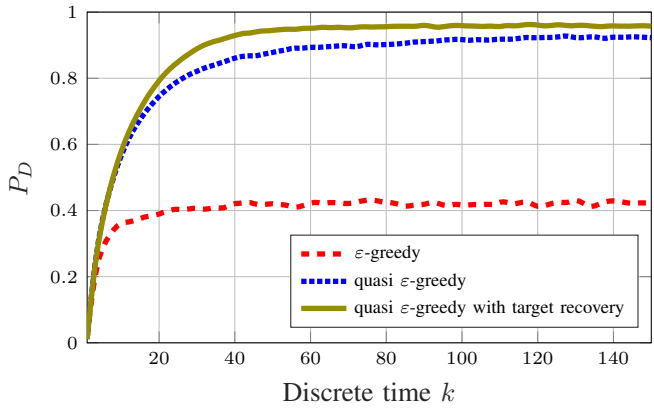}
  \caption{Policy comparison: $P_D$ of target 2 (Scenario 1).}
  \label{fig:Epsilon-greedy comparison target 2 (Scenario 1)}
\end{figure}

\cref{fig:Epsilon-greedy comparison target 2 (Scenario 1)} shows the $P_D$ of target 2 in Scenario 1 for the $\varepsilon$-greedy policy and the two proposed policies. The figure shows that both the quasi $\varepsilon$-greedy policy and quasi $\varepsilon$-greedy policy with target recovery have better detection performances than the classical $\varepsilon$-greedy one. In particular, the quasi $\varepsilon$-greedy policy with target recovery has the best performances among the three policies. 

\begin{figure}[t!]
  \centering
  \input{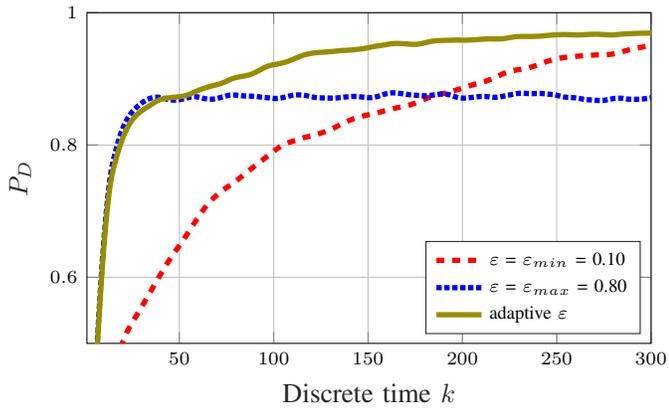}
  \caption{Adaptive vs static $\varepsilon$: $P_D$ of target 2 (Scenario 1).}
  \label{fig:varying epsilon: $P_D$ of target 2 (Scenario 1)}
\end{figure}

\begin{figure}[t!]
  \centering
%
%
\begin{tikzpicture}[trim axis left, trim axis right]
\tikzstyle{every node}=[font=\scriptsize]
\begin{axis}[%
width=75mm,
height=44mm,
at={(0\textwidth,0\textwidth)},
scale only axis,
xmin=1,
xmax=100,
xlabel style={font=\color{white!15!black}},
xlabel={Discrete time $k$},
ymin=0.5,
ymax=0.9,
ylabel style={font=\color{white!15!black}},
ylabel={$P_D$},
axis background/.style={fill=white},
xmajorgrids,
ymajorgrids,
legend style={at={(44mm,50mm)}, anchor=south west, legend cell align=left, align=left, draw=white!15!black, font=\small}
]
\addplot [color=red, dashed, line width=2.0pt]
  table[row sep=crcr]{%
1	0.092113463842405\\
2	0.139907819176601\\
3	0.185528150857165\\
4	0.229017841846486\\
5	0.270419542430109\\
6	0.309772750525235\\
7	0.347112913828575\\
8	0.382470733616559\\
9	0.415871280959549\\
10	0.447332784801023\\
11	0.476864855486276\\
12	0.50446679618389\\
13	0.530124945601419\\
14	0.553810590933646\\
15	0.575476850885076\\
16	0.59505430077074\\
17	0.612446652353758\\
18	0.62752890636663\\
19	0.64018573510682\\
20	0.650576972441985\\
21	0.65993849758132\\
22	0.669862117472787\\
23	0.679306640157608\\
24	0.688251566483796\\
25	0.696686310789731\\
26	0.70459929250024\\
27	0.711991131582694\\
28	0.718872695487878\\
29	0.725276254520315\\
30	0.731246861400215\\
31	0.736858423124575\\
32	0.742185297950059\\
33	0.747270281229624\\
34	0.752136526721486\\
35	0.756785821818687\\
36	0.761216688714662\\
37	0.765420726928393\\
38	0.769391192463328\\
39	0.773116708274989\\
40	0.776600801797805\\
41	0.779859799499259\\
42	0.782920696864289\\
43	0.785788615956879\\
44	0.788472068541866\\
45	0.790998615095498\\
46	0.793418314674988\\
47	0.795774038599372\\
48	0.798079280368142\\
49	0.800326055477674\\
50	0.802506784419022\\
51	0.804622256911374\\
52	0.80667070607056\\
53	0.808610977267202\\
54	0.810399013991309\\
55	0.812017826947585\\
56	0.81348986143677\\
57	0.814840720504753\\
58	0.816086372172922\\
59	0.81723415700567\\
60	0.818316531327444\\
61	0.819385790284786\\
62	0.820489304045942\\
63	0.821651335627989\\
64	0.822855222540891\\
65	0.82406089142345\\
66	0.825224424494636\\
67	0.82632182556546\\
68	0.827332378557492\\
69	0.828271869560841\\
70	0.829131646906712\\
71	0.829919677175335\\
72	0.83065946203526\\
73	0.831377112868782\\
74	0.832091426827363\\
75	0.832798008085253\\
76	0.8334855376506\\
77	0.834138215470895\\
78	0.834733901242765\\
79	0.835243163245883\\
80	0.835648170390955\\
81	0.835961154125952\\
82	0.836378126756138\\
83	0.836870789135286\\
84	0.83740563272064\\
85	0.837964533465651\\
86	0.838539335982385\\
87	0.839127365656281\\
88	0.839728208340055\\
89	0.840343408691769\\
90	0.840974816595903\\
91	0.84162432656865\\
92	0.842293708361688\\
93	0.842984589869181\\
94	0.843698107613262\\
95	0.844435217226664\\
96	0.845196514605852\\
97	0.845982386094809\\
98	0.846792943798349\\
99	0.847628035663429\\
100	0.848486979346591\\
};
\addlegendentry{$\alpha$ = $\alpha_{min}$ = 0.20}

\addplot [color=blue, densely dotted, line width=2.0pt]
  table[row sep=crcr]{%
1	0.060621424308635\\
2	0.125852911372936\\
3	0.187924814853793\\
4	0.246870255725747\\
5	0.30272386645233\\
6	0.355520758217507\\
7	0.405296332076803\\
8	0.452085160574561\\
9	0.495919235285508\\
10	0.536826737086934\\
11	0.574831215749231\\
12	0.609950212902524\\
13	0.642192063857613\\
14	0.671550529359263\\
15	0.697998161778034\\
16	0.721475931149806\\
17	0.741876202243619\\
18	0.759016577890746\\
19	0.772625072296726\\
20	0.782500720593634\\
21	0.789351349026836\\
22	0.795034613765569\\
23	0.799431802659483\\
24	0.802794884418193\\
25	0.805345349625673\\
26	0.807262692184928\\
27	0.80871497343103\\
28	0.809870562006027\\
29	0.810867802437687\\
30	0.811781337758656\\
31	0.812628288881236\\
32	0.813418825650032\\
33	0.81417690571943\\
34	0.814912606700632\\
35	0.815621772976135\\
36	0.816295772640846\\
37	0.816948939857012\\
38	0.817585037541509\\
39	0.818199534991194\\
40	0.818784059751034\\
41	0.819332039208727\\
42	0.819836151517246\\
43	0.82030771106332\\
44	0.820768839764415\\
45	0.821251276647484\\
46	0.821787365309096\\
47	0.822397150669054\\
48	0.82309424514142\\
49	0.82387352233239\\
50	0.824703399215826\\
51	0.825530218447128\\
52	0.826308534459119\\
53	0.827017557109512\\
54	0.827661671385679\\
55	0.828236063926313\\
56	0.828742413681879\\
57	0.82919414834239\\
58	0.829600715797766\\
59	0.829967233902221\\
60	0.830305195125906\\
61	0.830645160428222\\
62	0.831001471931202\\
63	0.83137074197381\\
64	0.831734889041689\\
65	0.832049406515768\\
66	0.832271716175769\\
67	0.832395645621391\\
68	0.83244263845662\\
69	0.832442236317562\\
70	0.832428419955069\\
71	0.832415149103284\\
72	0.832397933204417\\
73	0.832371965224659\\
74	0.832338258263403\\
75	0.832309709950861\\
76	0.832273473270393\\
77	0.832222948916762\\
78	0.832154679107335\\
79	0.832046509255674\\
80	0.831888881389188\\
81	0.831745212626368\\
82	0.831555845192157\\
83	0.831340289411615\\
84	0.831116493393277\\
85	0.830892343535032\\
86	0.83067094406361\\
87	0.830454045043126\\
88	0.830243087878542\\
89	0.83003915026266\\
90	0.829843017681751\\
91	0.829654721901995\\
92	0.829473617829642\\
93	0.829298716137537\\
94	0.829128932509382\\
95	0.828963485707002\\
96	0.828801702424787\\
97	0.828643326382789\\
98	0.828488626716136\\
99	0.828337646737715\\
100	0.828190911404523\\
};
\addlegendentry{$\alpha$ =$\alpha_{max}$ = 0.60}

\addplot [color=olive, solid, line width=2.0pt]
  table[row sep=crcr]{%
1	0.063494771766196\\
2	0.128629644461536\\
3	0.190638731167997\\
4	0.249557903215063\\
5	0.305424578550986\\
6	0.358276547824776\\
7	0.408151071650167\\
8	0.455083738930124\\
9	0.499107701319448\\
10	0.540252372508579\\
11	0.578541536870174\\
12	0.613990744555429\\
13	0.646604010539252\\
14	0.676369279686747\\
15	0.703252388541536\\
16	0.727187516108765\\
17	0.748060824517875\\
18	0.765687627896159\\
19	0.779804780663432\\
20	0.790252930642668\\
21	0.797841492688604\\
22	0.804472075139096\\
23	0.809816700055077\\
24	0.814103740407207\\
25	0.817527040147051\\
26	0.820262554750142\\
27	0.822474034166476\\
28	0.824291068072181\\
29	0.82581411482151\\
30	0.827133843372372\\
31	0.82833652537686\\
32	0.829481305283363\\
33	0.830601245482815\\
34	0.831703091214571\\
35	0.832782250076583\\
36	0.833836898097824\\
37	0.834872889982445\\
38	0.835890695428383\\
39	0.836890586554145\\
40	0.837858245923886\\
41	0.838765015506349\\
42	0.839596178691638\\
43	0.840338656471393\\
44	0.840992779640997\\
45	0.841543253845819\\
46	0.841958800010676\\
47	0.842213831257998\\
48	0.842311304534019\\
49	0.842265526876202\\
50	0.842117724484478\\
51	0.841933788151215\\
52	0.841762763482594\\
53	0.841626314555917\\
54	0.841527683728617\\
55	0.84147408859276\\
56	0.841466348032269\\
57	0.841524466831261\\
58	0.841677512248816\\
59	0.841926770865968\\
60	0.842243095774913\\
61	0.842569738623336\\
62	0.842846509143971\\
63	0.843020009677311\\
64	0.843057664381503\\
65	0.842941779889732\\
66	0.842691090948707\\
67	0.842355321633477\\
68	0.841987537108671\\
69	0.841634725098941\\
70	0.841332093551955\\
71	0.841105529288122\\
72	0.84096341841561\\
73	0.840886522998201\\
74	0.840842511764953\\
75	0.840810318618176\\
76	0.840769300762933\\
77	0.840714165647448\\
78	0.840643941053252\\
79	0.840574116873161\\
80	0.840548589479886\\
81	0.840640779637906\\
82	0.840937085802411\\
83	0.841326302628464\\
84	0.841752394150112\\
85	0.84219455127935\\
86	0.842646382302156\\
87	0.843107745499283\\
88	0.843580680486848\\
89	0.844068300154162\\
90	0.844574215602176\\
91	0.845101881730009\\
92	0.845654317893403\\
93	0.846233693217773\\
94	0.846841288765191\\
95	0.847477675232622\\
96	0.848142646273222\\
97	0.848835301214148\\
98	0.849554172928976\\
99	0.850297108189425\\
100	0.851061584170131\\
};
\addlegendentry{adaptive $\alpha$}

\end{axis}
\end{tikzpicture}%
  \caption{Adaptive vs static $\alpha$: $P_D$ of target 1 (Scenario 2).}
  \label{fig:varying alpha: $P_D$ of target 1 (Scenario 2)}
\end{figure}
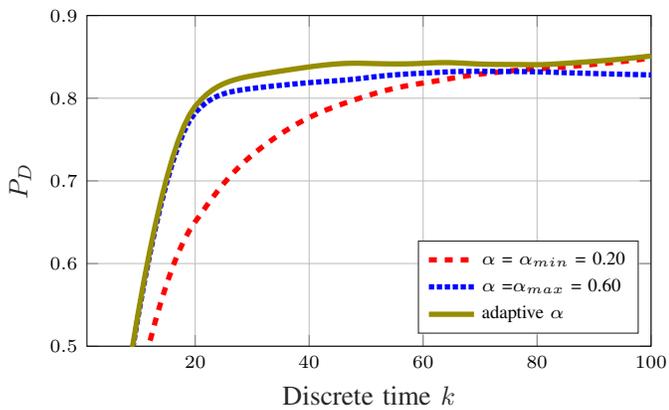

\cref{fig:varying epsilon: $P_D$ of target 2 (Scenario 1)} compares the performance of the adaptive $\varepsilon$ algorithm with the two static cases with $\varepsilon=\varepsilon_{min}$ and $\varepsilon=\varepsilon_{max}$ while keeping $\alpha$ constant ($\alpha=0.5$); \cref{fig:varying alpha: $P_D$ of target 1 (Scenario 2)} does the same for $\alpha$ when $\varepsilon$ is kept constant ($\varepsilon=0.5$). All the graphs show that the adaptive algorithm combines the positive effects of having high values of both the parameters in the initial transitory phase, when the system has to gather information about the surrounding environment, and low values in the following phase when the system approaches a steady state.

\begin{figure*}[t]
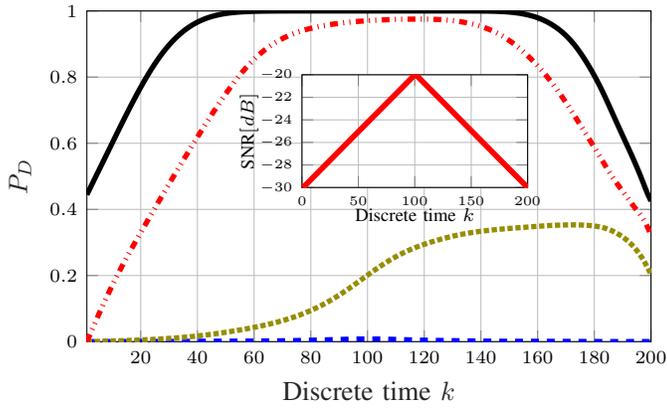
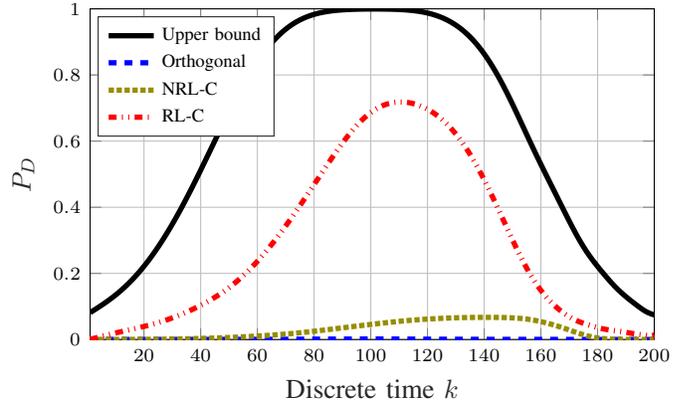

	\captionsetup[subfigure]{aboveskip=0pt,belowskip=-0pt}
	\centering
	\begin{subfigure}[t]{75mm}
   		\centering
    	\pgfmathsetlength{\imagewidth}{\linewidth}%
    	\pgfmathsetlength{\imagescale}{\imagewidth/524}%
    	\begin{tikzpicture}[x=\imagescale,y=-\imagescale]
        	\node[anchor=north west] at (0,0) {\input{figures/AlgorithmComparison/PerformanceComparison/Scenario5/Pd_Target1_Scenario5.tex}};
            \node[anchor=north west] at (200,60) {\input{figures/Appendix_SimulationFramework/SNR_Scenario5.tex}};
    	\end{tikzpicture}
  		\caption{target 1}
  		\label{fig:Performance comparison Scenario3: Pd target 1}	
	\end{subfigure}
	\hfill
	\begin{subfigure}[t]{75mm}
  		\centering
		\input{figures/AlgorithmComparison/PerformanceComparison/Scenario5/Pd_Target2_Scenario5.tex}
  		\caption{target 2}
  		\label{fig:Performance comparison Scenario3: Pd target 2}
	\end{subfigure}
	\caption{$P_D$ of the targets in Scenario 3. The \textcolor{red}{inset figure} in (a) shows the \textcolor{red}{SNR of both targets} expressed in dB.}
	\label{fig:Performance comparison Scenario 3}	
\end{figure*}

Scenario 3 consists of two targets with fixed angular position and variable SNR, which is shown in the inset in \cref{fig:Performance comparison Scenario3: Pd target 1}. \cref{fig:Performance comparison Scenario 3} shows the $P_D$ of both targets. Since the SNR increases in the interval [1,100] and decreases in [101,200], the $P_D$ varies accordingly. It is interesting to note that there is a delay between the drop of the $P_D$ and the time instant when the SNR starts to go down ($k=100$) for both the RL and NRL cognitive algorithms. For example, considering the performance of the RL-C algorithm in \cref{fig:Performance comparison Scenario3: Pd target 2}, the $P_D$ goes down after $k=110$; in the interval [101,110] the positive effect of the algorithm's learning capability prevails over the negative effect of the decreasing of the target's SNR. The RL-based algorithm shows better detection performance than the non RL-based cognitive one, confirming that the system is able to exploit the information gathered from the surrounding environment. Since the SNR of both targets is low, the orthogonal algorithm can't detect them. Even though the RL-based algorithm shows far better performance than the other algorithms, there's still a gap with the upper bound on the $P_D$. 

\begin{figure*}[t]
	\captionsetup[subfigure]{aboveskip=0pt,belowskip=-0pt}
	\centering
	\begin{subfigure}[t]{47mm}
  		\centering
		\input{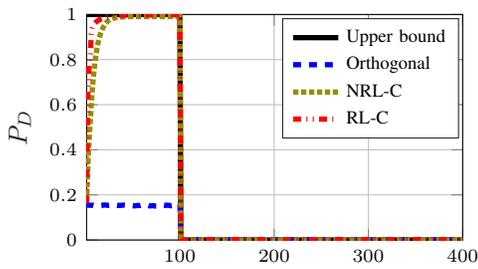}
  		\caption{target 1}
  		\label{fig:Performance comparison Scenario4: Pd target 1}
	\end{subfigure}
	\hfill
	\begin{subfigure}[t]{47mm}
  		\centering
		\input{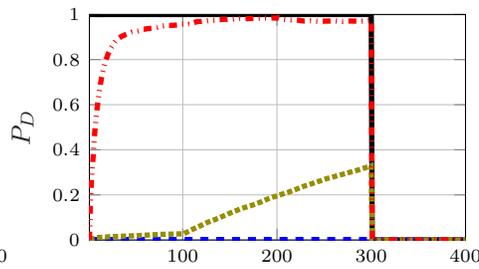}
  		\caption{target 2}
  		\label{fig:Performance comparison Scenario4: Pd target 2}	
	\end{subfigure}
	\hfill
	\begin{subfigure}[t]{47mm}
  		\centering
		\input{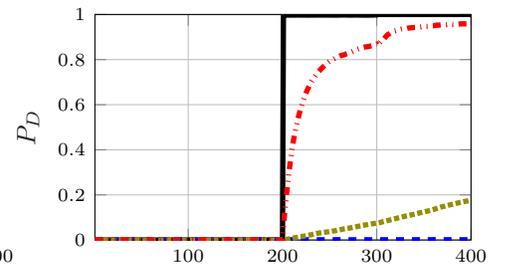}
  		\caption{target 3}
  		\label{fig:Performance comparison Scenario4: Pd target 3}
	\end{subfigure}
	\caption{$P_D$ of the targets in Scenario 4.}
	\label{fig:Performance comparison Scenario 4}	
\end{figure*}

\begin{figure}[t]
  \centering
  \input{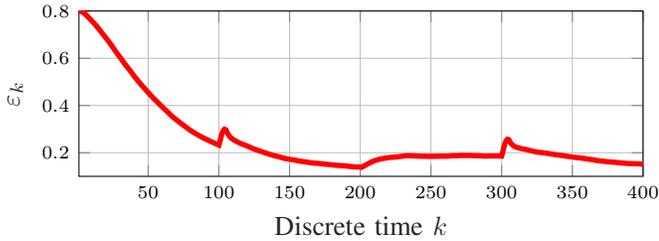}
  \caption{$\varepsilon_k$ sequence (Scenario 4)}
  \label{fig:Performance comparison Scenario4: epsilon}
\end{figure}

\begin{figure}[t]
  \centering
  \input{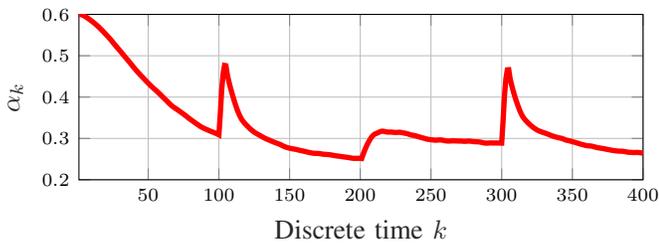}
  \caption{$\alpha_k$ sequence (Scenario 4)}
  \label{fig:Performance comparison Scenario4: alpha}
\end{figure}

Scenario 4 consists of three stationary targets. Target 1 and 2 are in the scene at time $k=1$ and disappear at time $k=101$ and $k=301$ respectively. Target 3 appears at time $k=201$ and stays in the scene until the end of the simulation. \cref{fig:Performance comparison Scenario 4} confirms that the RL-based algorithm is the one with the best detection performance among the three algorithms. \cref{fig:Performance comparison Scenario4: epsilon} and \cref{fig:Performance comparison Scenario4: alpha} show the temporal evolution of the $\varepsilon$ and $\alpha$ parameters: the adaptive algorithm is able to track the variations in the scene and adjust the parameters accordingly.

Finally, \cref{table:Pd Comparison with Ahmed} compares the performance of the proposed RL-C algorithm with one obtained in \cite{ahmed2021reinforcement}. Thanks to the developed quasi $\varepsilon$-greedy policy with target recovery and the adaptive selection of $\varepsilon$ and $\alpha$, a significant improvement has been achieved.

\begin{table}[t]
\caption{$P_D$ of the four targets in Scenario 3 in \cite{ahmed2021reinforcement} for a fixed $P_{FA}$ value of $10^{-4}$. The values in the second row are obtained using the proposed RL-C algorithm, while the ones in the third row are extracted from Figure 6 in \cite{ahmed2021reinforcement}.}
\label{table:Pd Comparison with Ahmed}
\centering
\def\arraystretch{1.2}
	\begin{tabular}{*{5}{|c}|}

 	\hline
 	Target & 1 ($\nu=-0.2$) & 2 ($\nu=0$) & 3 ($\nu=0.2$) & 4 ($\nu=0.3$) \\ 
 	\hline
 	RL-C & 1.00 & 0.98 & 0.99 & 0.97  \\ 
 	\hline
 	\cite{ahmed2021reinforcement} & 1.00 & 0.74 & 0.91 & 0.73  \\ 
	\hline
	\end{tabular}
\end{table}


\section{CONCLUSION}
\label{sec:CONCLUSION}

In the present paper we introduced two variations of the $\varepsilon$-greedy policy. In addition, we proposed an adaptive algorithm to select the SARSA $\varepsilon$ and $\alpha$ parameters, which increases both the RL-based algorithm performance and usability, since the user doesn't have to set them manually. Then the enhanced version of the algorithm proposed in \cite{ahmed2021reinforcement} was tested in various stationary and non-stationary scenarios. The results confirmed that the updated algorithm leads to a significant improvement of the detection capability of the massive MIMO radar system. On the other hand, the new policies and the adaptive selection of the hyper-parameters slightly increase the computational overhead, which must be taken into account when the algorithm is implemented by a real-time system. Furthermore, the time complexity of the algorithm grows exponentially with $N_T$, due to the growth of the dimension of the weighting matrix $W$ \cite{ahmed2021reinforcement}, limiting its scalability.
Future works will investigate the possibility to fill the gap between the performance of the improved RL-based algorithm and the upper bound. More specifically, we will try to fuse the position information (i.e. target tracking) with the RL-based detection algorithm.


\appendices

\section{SIMULATION FRAMEWORK DESCRIPTION}
\label{appendix}

In all the simulated scenarios the noise process is an AR(6) process with t-distributed independent and identically distributed (i.i.d.) innovations with the same parameters as the ones in \cite{ahmed2021reinforcement}. 

\cref{table:Target scenarios} contains all the details that characterise the four simulated scenarios.

\begin{table}[h!]
\centering
\def\arraystretch{1.2}
\caption{Target scenarios.}
\label{table:Target scenarios}
	\begin{tabular}{| c | >{\centering\arraybackslash} p{11mm} | c | >{\centering\arraybackslash} p{9mm} | c | c |} 
 	\hline
 	\multirow{2}{*}{Scenario} & Time & \multirow{2}{*}{Target} & Angular & \multirow{2}{*}{$\nu$}  & \multirow{2}{*}{$SNR_{dB}$}\\ 
 	& Interval & & Bin & &\\
 	
 	\hline 	
 	\rowcolor{Gray}
 	&& 1 & 7 & -0.20 & -20 \\  \cline{3-6} \rowcolor{Gray}
  	\multirow{-2}{*}{1} & \multirow{-2}{*}{[1,300]} & 2 & 16 & 0.25 & -20 \\
  	
	\hline
 	2 & [1,100] & 1 & 17 & 0.30 & -20\\
 	 
 	\hline
 	\rowcolor{Gray}
 	&& 1 & 7 & -0.20  & variable \\ \cline{3-5} \rowcolor{Gray}
 	\multirow{-2}{*}{3} & \multirow{-2}{*}{[1,200]} & 2 & 16 & 0.25 & (Inset Fig.\ref{fig:Performance comparison Scenario3: Pd target 1}) \\

 	\hline
 	\multirow{6}{*}{4} & \multirow{2}{*}{[1,100]} & 1 & 5 & -0.30 & -18 \\ \cline{3-6}
 	& & 2 & 13 & 0.10 & -21 \\
 	\cline{2-6}
	&[101,200] & 2 & 13 & 0.10 & -21 \\
 	\cline{2-6}
 	& \multirow{2}{*}{[201,300]} & 2 & 13 & 0.10 & -21 \\ \cline{3-6}
 	&& 3 & 17 & 0.30 & -20 \\
	\cline{2-6}
	& [301,400] & 3 & 17 & 0.30 & -20 \\
 	\hline
	\end{tabular}
\end{table}



\bibliography{biblio}
\bibliographystyle{ieeetr}

\correspauthor%


\clearpage

\end{document}